\documentclass{article}
\usepackage{spconf,amsmath,amssymb,graphicx}
\usepackage[utf8]{inputenc}
\usepackage[T1]{fontenc}
\usepackage{booktabs}
\usepackage{multirow}
\usepackage{xcolor}
\usepackage{float}
\usepackage[hidelinks]{hyperref}
\usepackage{url}
\usepackage{hyperref}

\newcommand{\fire}{\raisebox{-0.2em}{\includegraphics[height=1em]{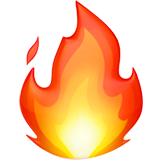}}}
\newcommand{\snow}{\raisebox{-0.2em}{\includegraphics[height=1em]{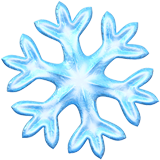}}}
\newcommand{\rel}[1]{\,{\scriptsize(${-}#1$)}}

\title{DASHA: Diarization-Aware Speech-LLM ASR for Healthcare in Code-Switched Indic Doctor-Patient Conversations}

\name{\begin{tabular}{@{}c@{}}S\'everin Baroudi$^{1,2}$, Yanis Labrak$^{2}$, Shashi Kumar$^{2,7}$, Joonas Kalda$^{3}$, Sergio Burdisso$^{2}$,\\
Pawel Cyrta$^{6}$, Juan Ignacio Alvarez-Trejos$^{3}$, Petr Motl\'{i}\v{c}ek$^{2}$, Herv\'e Bredin$^{3}$, Ricard Marxer$^{1,4,5}$\end{tabular}}
\address{\begin{tabular}{@{}c@{}}
$^1$Universit\'e de Toulon, Aix Marseille Univ, CNRS, LIS, France \quad
$^2$Idiap Research Institute, Switzerland \\
$^3$pyannoteAI, France \quad
$^4$Mila, Canada \quad
$^5$CNRS, ILLS, France \quad
$^6$Stenograf \quad
$^7$EPFL, Switzerland
\end{tabular}}

\begin{document}
\ninept
\renewcommand{\baselinestretch}{0.91}\selectfont
\maketitle

\begin{abstract}
Extracting patients' medical conditions from recorded primary-care conversations requires knowing who said what, which is difficult when speech is far-field, spontaneous, overlapping and code-switched. We study this problem on DISPLACE-M, a corpus of Hindi--English (Hinglish) doctor--patient conversations, and propose DASHA, a modular cascade of diarization, speaker-attributed ASR (SA-ASR) and LLM-based extraction. An end-to-end neural diarization with vector clustering (EEND-VC) system built on a multilingual w2v-BERT~2.0 encoder reduces DER from 9.31\% to 7.76\%. A Qwen3-ASR model adapted to Hindi and the target domain, with Devanagari normalization, reduces tcpWER from 26.78\% to 19.27\%, and to 18.59\% with optional LLM correction. A module-swap analysis shows that better diarization improves extraction only when paired with the adapted ASR. On code-switched words, the multilingual encoder roughly halves speaker confusion. DASHA sets a new state of the art for two-speaker, code-switched Hindi SA-ASR and ranked first among 25 participants in the DISPLACE-M challenge.
\end{abstract}


\section{Introduction}
\label{sec:intro}

\begin{figure}[t]
  \centering
  \includegraphics[width=0.92\columnwidth, trim=0.5cm 0.3cm 0 0, clip]{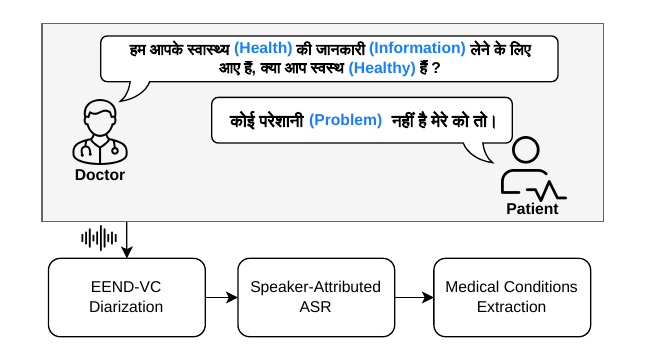}
  \vspace{-0.8em}
  \caption{A code-switched Hinglish exchange between a doctor and a patient, and the proposed cascade. English words, written in Devanagari in the references, are shown in blue with their English form.}
  \label{fig:convo}
  \vspace{-1.2em}
\end{figure}

Automatic documentation of clinical encounters has progressed rapidly, but most work targets monolingual English interactions~\cite{chiu18_interspeech, giorgi-etal-2023-wanglab} and often relies on large proprietary datasets.
Community healthcare in rural India is a much harder setting, conversations are recorded in the field with a single distant microphone, turns are short and often overlap, and speakers freely mix Hindi and English (Hinglish)~\cite{dey-fung-2014-hindi}, as in Fig.~\ref{fig:convo}.
In this setting, extracting what the patient reports requires knowing \emph{who} said \emph{what}, which is a multi-talker, speaker-attributed ASR (SA-ASR) problem.

Role-aware systems for professional conversations assign roles with speaker embeddings on top of ASR segments~\cite{kim25k_interspeech}. This assumes that speakers take turns one at a time, which rarely holds in spontaneous interaction.
Multi-talker and diarization-conditioned ASR~\cite{polok2026dicow} handle overlap more directly, but they have not been studied on low-resource, code-switched (CS) medical speech.
Moreover, standard metrics such as DER and tcpWER~\cite{cornell23_chime} average errors over whole conversations and do not show whether systems fail specifically at language switches.
CS-specific ASR metrics have been proposed~\cite{hamed2022benchmarking, ugan2025pier}, but, to our knowledge, they have not been applied jointly to diarization and recognition.

\begin{figure*}[t]
  \centering
  \includegraphics[width=\textwidth, trim=0.6cm 0.19cm 0.6cm 0.0cm, clip]{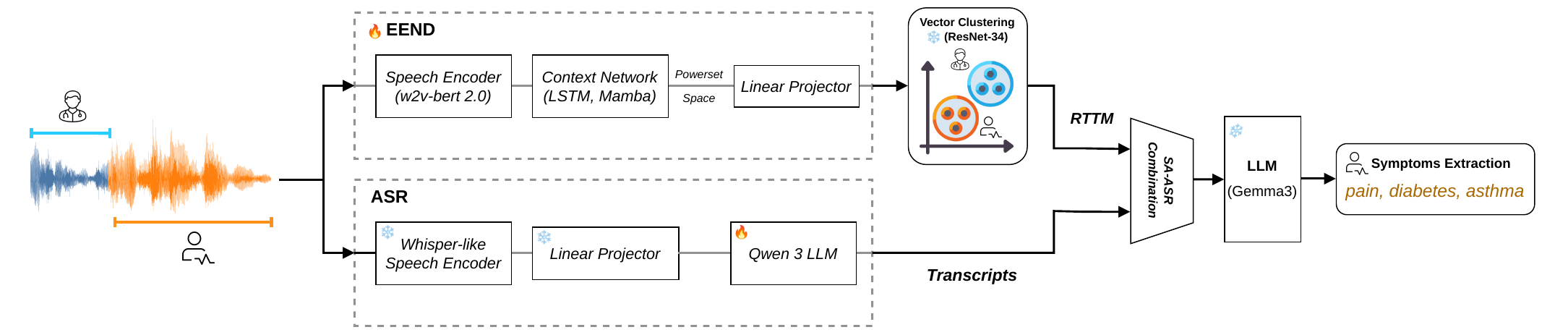}
  \vspace{-1.6em}
  \caption{Proposed DASHA framework: EEND-VC diarization $\rightarrow$ diarization-driven SA-ASR $\rightarrow$ LLM-based medical condition extraction. \fire{} and \snow{} denote fine-tuned and frozen modules, respectively.}
  \label{fig:pipeline}
  \vspace{-1.0em}
\end{figure*}

We address these gaps on the DISPLACE-M challenge~\cite{displace_m} with DASHA, a cascade of diarization, SA-ASR and extraction (Fig.~\ref{fig:pipeline}), in which each module can be replaced independently.
Most components are established techniques.

Our contributions are their adaptation to this setting and an analysis of how they interact:
(1)~an EEND-VC diarization system~\cite{eend-vc} with a multilingual w2v-BERT~2.0 encoder, together with a study of pre-training data, context network and output size;
(2)~an incremental adaptation study of Qwen3-ASR for Hinglish SA-ASR, which yields a 28\% relative tcpWER reduction with open models only and a new state of the art for two-speaker, code-switched Hindi SA-ASR;
(3)~a comparison of open and proprietary LLMs for condition extraction, from transcripts and directly from audio, and a module-swap analysis showing that diarization and ASR improvements are complementary;
(4)~a diagnostic evaluation restricted to code-switched words, which extends the point-of-interest idea of~\cite{ugan2025pier} to diarization.
Our submission ranked first among 25 participants in the challenge\footnote{\url{displace2026.github.io}}.
Code and model weights will be publicly released\footnote{\url{dasha-diarization.github.io}}.

\section{Data and protocol}
\label{sec:data}

DISPLACE-M~\cite{displace_m} contains de-identified conversations between doctors and patients, recorded in primary-care settings in rural India.
We merge the two released development sets (DEV1 and DEV2) into a single DEV set: 125 dialogues, 24.5\,h, and 30.1k turns (between 43 and 565 turns per dialogue; about 2.9\,s per turn on average).
In DEV, 84.30\% of the audio contains a single speaker, 11.35\% is silence and 4.35\% is overlapped speech.
The overlap ratio is moderate, but it occurs at frequent, short turn changes.
For in-domain adaptation, we train on most of DEV and hold out a 3\,h \emph{valid} split, which is used for early stopping and hyper-parameter selection.
All systems are evaluated on the blind evaluation set EVAL1 (10.1\,h), scored by the official challenge platform.
Following the challenge, DER is computed with overlap and without a forgiveness collar; SA-ASR is scored with tcpWER~\cite{cornell23_chime}, which aligns speaker-attributed words under the optimal speaker permutation and a temporal constraint.
Condition extraction is scored with ROUGE-1 and ROUGE-L F1 between predicted and reference condition lists.

\section{Proposed system}
\label{sec:system}

\subsection{Speaker diarization}
\label{ssec:diar}

Inspired by the architecture described in \cite{specializing} which have shown strong diarization performance on low-resouce languages, our starting point is the use of w2v-BERT~2.0~\cite{w2vbert} (585M parameters, pre-trained on 4.5M hours in 143 languages), as a robust multilingual backbone. The encoder is adapted with LoRA (rank 32), and fed into a context network, for which we compare 4 LSTM layers (2.4M parameters) with 7 bidirectional Mamba layers (7.4M)~\cite{mamba}, followed by a linear powerset classifier~\cite{Plaquet2023}. 
Since each dialogue involves one doctor and one patient, we compare a 2-speaker output layer with the usual 4-speaker one. Local segmentations are linked across chunks by clustering ResNet-34 speaker embeddings~\cite{wespeaker} extracted from single-speaker frames, as traditionally done in EEND-VC approaches. We use $k$-means clustering with $k{=}2$ instead of AHC or VBx, whose thresholds are hard to tune on little in-domain data. The overall diarization system is depicted in Figure \ref{fig:pipeline}.
We compare our approach to the commonly used open-source DiariZen~\cite{han2025leveraging} system. The latter leverages a WavLM-Base encoder~\cite{wavlm} trained mostly on English read speech, followed by a Conformer.

\vspace{1mm}

\noindent\textbf{Training.} Models are pre-trained on two different manually assembled diarization corpora.
\textbf{C1} (212\,h) contains DIHARD3~\cite{DIHARD}, VoxConverse~\cite{voxconverse}, MSDWild~\cite{MSDWild} and DISPLACE 2024~\cite{displace24}, while
\textbf{C2} (908\,h) adds the following meeting and conversational corpora AMI~\cite{AMI}, AliMeeting~\cite{alimeeting}, AISHELL-4~\cite{aishell4}, NOTSOFAR-1~\cite{notsofar} and RAMC~\cite{RAMC} to C1.
Some pre-trained model are then fine-tuned (FT) on the DISPLACE-M training split (as displayed in Table \ref{tab:diar}).
We use 10\,s chunks, a batch size of 32, and learning rates of $10^{-5}$, and $10^{-3}$ for the SSL part and context network, with bf16 precision on H100 GPUs.

\subsection{Speaker-attributed ASR}
\label{ssec:asr}

The diarization output defines speaker-homogeneous segments, which are transcribed independently and labeled with the corresponding speaker, as in Fig.~\ref{fig:pipeline}.
Transcribing only diarized speech also avoids transcribing background talkers and noise.
We use Qwen3-ASR-1.7B~\cite{shi2026qwen3asrtechnicalreport}, which combines a 300M-parameter AuT audio encoder with a Qwen3-1.7B decoder through a projector.
Diarization-labeled segments that are shorter than 0.4\,s are discarded during decoding because they mostly represent non-verbal vocalizations. 
The model is first fine-tuned on about 1,800\,h of public Hindi speech (FLEURS~\cite{fleurs2022arxiv}, IndicTTS~\cite{indictts2023}, IndicVoices~\cite{DBLP:conf/acl/JavedNGJBMSAFPR24}, VAANI~\cite{vaani2025}).
We then average the weights of several checkpoints and fine-tune the model on the DISPLACE-M training split.
Punctuation and words are normalized into canonical Unicode, as Devanagari characters can be encoded in several equivalent ways, for both training targets and hypotheses.
As an optional final step, we apply dialogue-level generative error correction (GEC)~\cite{chen2023hyporadise}. An LLM is prompted to make only minimal, high-confidence edits that preserve segmentation and code-mixing. It receives three contrastive examples: ASR/reference excerpts around each error, with two segments of context on either side, which worked better than full-dialogue examples.
Among the LLMs we tried on the valid split, GPT-4.1 worked best. Because it is proprietary, we report results with and without this step.

We compare our approach to the current state of the art for Hindi dialect ASR systems, IndicConformer-600M\hyperlink{fn:indic}{$^\dagger$}, a hybrid CTC/RNN-T model covering the 22 official languages of India.

\subsection{Medical condition extraction}
\label{ssec:extraction}

In a cascade, each module works only on the output of the one before it. Therefore, diarization errors (misattributed turns) and ASR errors (misrecognized medical or code-switched terms) pass on to the extractor, which cannot recover information lost upstream.
An end-to-end (E2E) model that extracts conditions directly from the audio avoids this compounding, but gives up modularity.
We evaluate both paradigms.

\vspace{1mm}

\noindent\textbf{Cascade.} An LLM extracts the patient's conditions from our Hinglish SA-ASR transcripts, using in-context learning (ICL) with 0 or 6 examples drawn from the training split.
We compare open models (Gemma~3 4B/12B/27B~\cite{gemma_2025}) with proprietary ones (Gemini~3 Pro, Claude Opus~4.1, Amazon Nova Pro v1, all accessed through their official APIs).

\vspace{1mm}

\noindent\textbf{E2E.} An audio-capable LLM receives the recording and the extraction instructions, with no intermediate transcript.
We consider open omni-modal models (Kimi-Audio 7B, Qwen2.5-Omni 3B/7B) and the proprietary Gemini~3 Pro.
For Gemini~3 Pro, we also test two prompt variants that add either our SA-ASR transcript or the list of admissible conditions to the audio input.

\section{Results}
\label{sec:results}

\subsection{Speaker diarization}
\label{ssec:res-diar}

Table~\ref{tab:diar} lists the diarization systems (D1--D7).
Our custom diarization system (D1) outperforms DiariZen, thanks to the multilingual w2v-BERT~2.0 encoder, reducing DER from 9.31\% to 8.52\% (8.5\% relative).
Pre-training on the larger C2 compound is slightly worse (D2: 8.96\%), adding 700\,h of mostly meeting-style speech does not help in this far-field clinical setting. The added corpora are mostly multi-party meetings, which do not match these two-speaker recordings on in-the-wild settings.
In-domain fine-tuning gives the largest gain (0.8--1.1 points absolute), and D3 obtains the best DER of 7.76\% (16.6\% relative to DiariZen).
The fine-tuned variants (D3--D7) lie within 0.11 points of each other. Given these small differences, we interestingly conclude that neither the Mamba context network nor the 2-speaker output gave a measurable benefit over the LSTM with 4 speaker outputs.
We use D3 in the rest of the paper.

\begin{table}[H]
  \centering
  \caption{Diarization results on EVAL1 test set.}
  \label{tab:diar}
  \vspace{1mm}
  \setlength{\tabcolsep}{3.5pt}
  \begin{tabular}{l c c c c c}
    \toprule
    \textbf{ID} & \textbf{Pre-train} & \textbf{Context} & \textbf{\#Spk out} & \textbf{FT} & \textbf{DER (\%)}$\downarrow$ \\
    \midrule
    DiariZen\hyperlink{fn:diarizen}{$^{\ast}$} & -- & Conformer & 4 & -- & 9.31 \\
    \midrule
    D1 & C1 & LSTM  & 4 & -- & 8.52 \\
    D2 & C2 & LSTM  & 4 & -- & 8.96 \\
    \midrule
    D3 & C1 & LSTM  & 4 & \checkmark & \textbf{7.76} \\
    D4 & C2 & LSTM  & 4 & \checkmark & 7.84 \\
    D5 & C2 & Mamba & 4 & \checkmark & 7.87 \\
    D6 & C2 & LSTM  & 2 & \checkmark & 7.81 \\
    D7 & C2 & Mamba & 2 & \checkmark & 7.81 \\
    \bottomrule
  \end{tabular}
\end{table}
{\renewcommand{\thefootnote}{$\ast$}%
 \footnotetext{\hypertarget{fn:diarizen}{}\href{https://huggingface.co/BUT-FIT/diarizen-wavlm-base-s80-md}{\nolinkurl{hf.co/BUT-FIT/diarizen-wavlm-base-s80-md}}}}

\subsection{Speaker-attributed ASR}
\label{ssec:res-asr}

Table~\ref{tab:asr} reports the incremental ASR study. All Qwen3 rows use D3 diarization, so that differences come from the ASR module only. IndicConformer reaches 26.78\% tcpWER with DiariZen annotations, and 26.24\% with our D3 diarization outputs.
In zero-shot, Qwen3-ASR is slightly worse (27.66\%), which reflects the mismatch between its generic training data and noisy, spontaneous Hinglish.
Fine-tuning on public Hindi speech (A1) and checkpoint averaging (A2) close this gap.
Fine-tuning on in-domain data (A3) gives by far the largest gain ($-6.1$ points). This suggests that the domain (acoustic conditions, turn-taking, medical and code-switched vocabulary) matters more than the amount of generic Hindi data.
Unicode normalization (A4) removes spurious mismatches between visually identical Devanagari strings.
A4 is our best fully open system, with a 28.0\% relative reduction over the reference (26.78\%$\rightarrow$19.27\%).
The optional proprietary GEC pass (A5) gives a further 0.7-point reduction, to 18.59\%.
Since the GEC may in principle add or remove content, it should be used with human review in clinical applications.

\begin{table}[H]
  \centering
  \caption{Incremental SA-ASR results on EVAL1 (tcpWER, \%). All rows below the first use D3 diarization.}
  \label{tab:asr}
  \vspace{1mm}
  \setlength{\tabcolsep}{4pt}
  \begin{tabular}{l l r}
    \toprule
    \textbf{ID} & \textbf{System} & \textbf{tcpWER}$\downarrow$ \\
    \midrule
     & IndicConformer-600M\hyperlink{fn:indic}{$^\dagger$} + DiariZen & 26.78 \\
      & IndicConformer-600M\hyperlink{fn:indic}{$^\dagger$} + D3 diarization & 26.24 \\
    \midrule
     & Qwen3-ASR-1.7B (zero-shot) & 27.66 \\
    A1 & \quad + FT on public Hindi data & 26.02 \\
    A2 & \quad + checkpoint averaging & 25.73 \\
    A3 & \quad + FT on DISPLACE-M & 19.61 \\
    A4 & \quad + Unicode/punctuation normalization & \textbf{19.27} \\
    A5 & \quad + dialogue-level GEC$^\dagger$ & 18.59 \\
    \bottomrule
  \end{tabular}
\end{table}
{\renewcommand{\thefootnote}{$^\dagger$}%
 \footnotetext{\hypertarget{fn:indic}{}\href{https://huggingface.co/ai4bharat/indic-conformer-600m-multilingual}{\nolinkurl{hf.co/ai4bharat/indic-conformer-600m-multilingual}}}}

\subsection{Medical condition extraction}
\label{ssec:res-extract}

We first compare E2E models on DEV in zero-shot (Table~\ref{tab:omni}).
Kimi-Audio mostly produces hallucinated output, and the best open model, Qwen2.5-Omni 7B, reaches 30.9 ROUGE-1, 3.6 points below Gemini~3 Pro.
We therefore use Gemini~3 Pro as the E2E system in the remaining experiments.

\begin{table}[H]
  \centering
  \caption{Zero-shot E2E extraction with omni-modal models on DEV (ROUGE F1, \%).}
  \label{tab:omni}
  \vspace{1mm}
  \small
  \begin{tabular}{l c c c}
    \toprule
    \textbf{Model} & \textbf{Size} & \textbf{R-1}$\uparrow$ & \textbf{R-L}$\uparrow$ \\
    \midrule
    Kimi-Audio & 7B   & 0.03 & 0.03 \\
    \midrule
\multirow{2}{*}{Qwen2.5-Omni} & 3B & 25.3 & 21.7 \\
                              & 7B & 30.9 & 27.2 \\
    \midrule
    Gemini 3 Pro & --   & \textbf{34.5} & \textbf{34.5} \\
    \bottomrule
  \end{tabular}
\end{table}

In the cascade (Table~\ref{tab:extract}), 6-shot prompting improves every model.
Scale does not help monotonically, Gemma~3 12B outperforms the 27B model in both settings.
The best open model (28.97 ROUGE-1) remains about 6 points behind the best proprietary one (Claude Opus~4.1, 34.90).
Prompting Gemini~3 Pro directly on audio gives the highest scores (45.60 ROUGE-1, zero-shot).
Here, 6-shot examples \emph{reduce} performance (38.78).
Adding our transcript to the audio input does not help in zero-shot but partly offsets the 6-shot drop. Providing the list of admissible conditions hurts most (45.60$\rightarrow$31.68 ROUGE-1). The model likely copies list entries given in the prompt, which ROUGE penalizes.
These results are consistent with the hypothesis that bypassing the transcription bottleneck from the cascaded approach, with an E2E one, allows for better extraction.
The cascade approach remains attractive when data cannot leave the premises, because every module can then be run locally with open models.

\begin{table}[H]
  \centering
  \caption{Medical condition extraction on EVAL1 (ROUGE F1, \%). Cascade rows use D3+A5 transcripts.}
  \label{tab:extract}
  \vspace{1mm}
  \setlength{\tabcolsep}{5.5pt}
  \begin{tabular}{l cc cc}
    \toprule
    & \multicolumn{2}{c}{\textbf{Zero-shot}} & \multicolumn{2}{c}{\textbf{6-shot}} \\
    \cmidrule(lr){2-3}\cmidrule(lr){4-5}
    \textbf{Model} & \textbf{R-1}$\uparrow$ & \textbf{R-L}$\uparrow$ & \textbf{R-1}$\uparrow$ & \textbf{R-L}$\uparrow$ \\
    \midrule
    \multicolumn{5}{l}{\textit{Open-source Cascade (text input)}} \\
    \;\;\;Gemma 3 4B  & 15.46 & 13.87 & 18.27 & 15.62 \\
    \;\;\;Gemma 3 12B & 21.79 & 19.08 & 28.97 & 26.62 \\
    \;\;\;Gemma 3 27B & 19.22 & 17.21 & 25.25 & 24.50 \\
    \midrule
    \multicolumn{5}{l}{\textit{Closed-source Cascade (text input)}} \\
    \;\;\;Gemini 3 Pro        & 28.60 & 25.52 & 32.36 & 30.72 \\
    \;\;\;Claude Opus 4.1     & 25.75 & 24.34 & \textbf{34.90} & \textbf{32.72} \\
    \;\;\;Nova Pro v1         & --    & --    & 31.16 & 29.21 \\
    \midrule
    \multicolumn{5}{l}{\textit{Closed-source E2E (Gemini 3 Pro, audio input)}} \\
    \;\;\;Audio               & \textbf{45.60} & \textbf{43.55} & 38.78 & 38.29 \\
    \;\;\;\;\;\; + Transcript  & 43.62 & 42.07 & 41.61 & 39.73 \\
    \;\;\;\;\;\; + Condition list & 31.68 & 31.05 & -- & -- \\
    \bottomrule
  \end{tabular}
\end{table}

\subsection{Interaction between modules}
\label{ssec:res-synergy}

Table~\ref{tab:synergy} swaps the diarization and ASR modules while keeping the extractor fixed.
Upgrading only the ASR improves ROUGE-1 from 25.16 to 27.90.
Upgrading only the diarization lowers it to 23.72, even though D3 slightly improves tcpWER with the same ASR (Table~\ref{tab:asr}).
Upgrading both gives the best result (28.97, +15.1\% relative).
Transcript quality therefore dominates downstream performance, and the diarization gains translate into extraction gains only with the adapted ASR.
The overall results shows that modules of a cascade should be evaluated together, not only in isolation.

\begin{table}[H]
  \centering
  \caption{Module-swap analysis: 6-shot Gemma~3 12B extraction on EVAL1 with baseline or proposed diarization and ASR.}
  \label{tab:synergy}
  \vspace{1mm}
  \begin{tabular}{l l c c}
    \toprule
    \textbf{ASR} & \textbf{Diarization} & \textbf{R-1}$\uparrow$ & \textbf{R-L}$\uparrow$ \\
    \midrule
    \multirow{2}{*}{IndicConformer} & DiariZen & 25.16 & 22.91 \\
                                    & D3             & 23.72 & 20.24 \\
    \midrule
    \multirow{2}{*}{A5}             & DiariZen & 27.90 & 25.79 \\
                                    & D3             & \textbf{28.97} & \textbf{26.62} \\
    \bottomrule
  \end{tabular}
\end{table}


\section{Analysis of code-switched segments}
\label{sec:cs}

\noindent\textbf{Protocol.}
We ask whether the gains in Sec.~\ref{sec:results} also hold on code-switched (English-origin) words.
Three LLMs (Claude Haiku~4.5, Gemini~2.5 Flash, GPT-5.4 mini) were tasked to independently label the DEV vocabulary, isolating the CS words on which they agreed. The resulting list was then manually checked, yielding a CS lexicon of 372 words.
DEV provides no word timings, so we trained a Hindi Montreal Forced Aligner~\cite{MFA} model on the DEV transcripts and aligned all non-overlapped speech.
The CS intervals cover 5.55\% of the speech and contain 1,517 tokens (4.65\% of all words).
CS-DER is the DER restricted to these intervals (no collar).
CS-WER counts ASR errors only on CS reference tokens. It is the point-of-interest error rate (PIER)~\cite{ugan2025pier} with CS words as points of interest.
The analysis is run on DEV, comparing the impact of  finetuned (D3 and A5) and non-finetuned versions (DiariZen, IndicConformer, D1 and A1) on CS portions.
Results for D3 and A5 (marked $^\ddagger$) are given for reference only.


\begin{table}[H]
  \centering
  \caption{Diagnostic evaluation on CS words (DEV). Relative reductions w.r.t.\ the baseline in parentheses.}
  \label{tab:cs}
  \vspace{1mm}
  \setlength{\tabcolsep}{2.6pt}
  \resizebox{\columnwidth}{!}{%
  \begin{tabular}{l|cc|ccc}
    \toprule
    \multicolumn{6}{c}{\textit{Diarization (\%)}} \vspace{1mm} \\ 
    \textbf{System} & \textbf{DER}$\downarrow$ & \textbf{CS-DER}$\downarrow$ & \textbf{Miss} & \textbf{FA} & \textbf{Conf.} \\
    \midrule
    \;\;DiariZen & 12.21 & 3.38 & 0.69 & 1.07 & 1.62 \\
    \;\;D1 & 10.49\rel{14\%} & 2.69\rel{20\%} & 0.55 & 1.25 & 0.89\rel{45\%} \\
    \;\;D3 & 8.47\rel{31\%} & 2.54\rel{25\%} & 0.59 & 1.14 & 0.81\rel{50\%} \\
    \midrule
    \midrule
    \multicolumn{6}{c}{\textit{ASR (\%)}} \vspace{1mm}  \\
    \textbf{System} & \textbf{WER}$\downarrow$ & \textbf{CS-WER}$\downarrow$ & \textbf{Ins} & \textbf{Del} & \textbf{Sub} \\
    \midrule
    \;\;IndicConformer & 25.52 & 28.19 & 1.12 & 4.72 & 22.35 \\
    \;\;A1 & 26.04 & 29.12 & 1.57 & 2.10 & 25.45 \\
    \;\;A5 & 19.10\rel{25\%} & 21.58\rel{23\%} & 1.16 & 2.14 & 18.28\rel{18\%} \\
    \bottomrule
  \end{tabular}%
  }
\end{table}

\begin{figure}[H]
  \centering
  \includegraphics[width=\columnwidth]{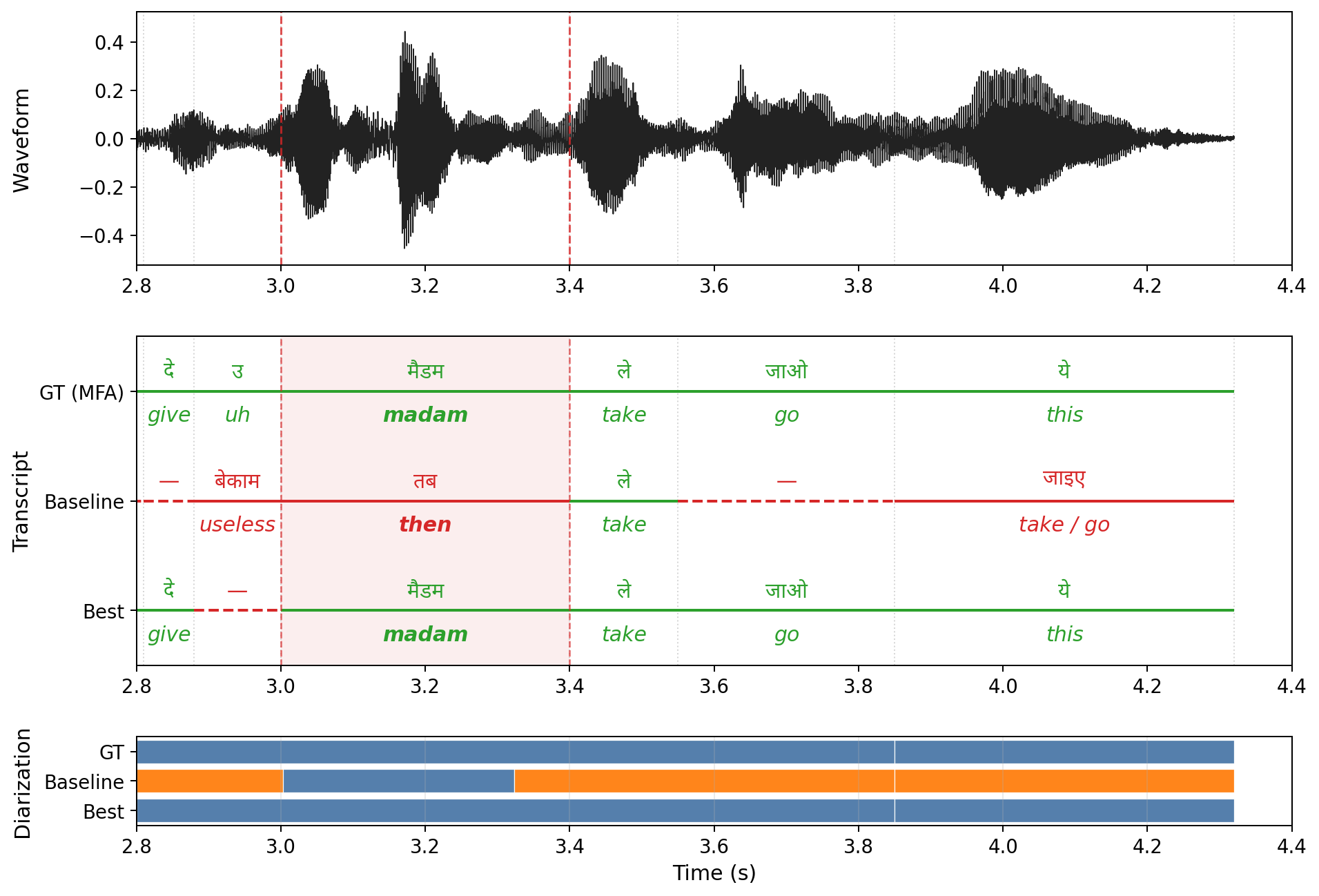}
  \caption{Example around the word \emph{madam} (a CS or borrowed word; Hindi script: \emph{mai\d{d}am}). Middle: MFA-aligned reference and hypotheses of the baseline and the best systems (errors in red). Bottom: diarization, with one color per speaker.}
  \label{fig:example}
\end{figure}

\noindent\textbf{Diarization.}
In the clean comparison (Table~\ref{tab:cs}), the w2v-BERT~2.0 encoder (D1) reduces CS-DER by 20\% relative, compared with 14\% for the overall DER.
This reduction comes from speaker confusion, which drops by 45\% (1.62$\rightarrow$0.89), while false alarms slightly increase. As CS intervals contain only single-speaker speech, this means D1 more often hypothesizes a spurious second speaker there.
In-domain fine-tuning (D3) further lowers confusion only slightly (0.89$\rightarrow$0.81), and its CS and overall gains are similar in size (25\% vs.\ 31\%). 
Fig.~\ref{fig:example} illustrates the typical baseline error: a sudden speaker change inside a continuous turn, around a CS word.
CS intervals exclude overlapped speech by construction, so absolute CS-DER values should not be compared with the overall DER, which includes overlap.

\vspace{1mm}

\noindent\textbf{ASR.}
CS-WER is higher than WER for every system (by about 2.5--3 points), so CS words are harder to recognize than average.
Fine-tuning on public Hindi data only (A1) halves deletions on CS words but increases substitutions (22.35$\rightarrow$25.45), so CS-WER gets worse (28.19$\rightarrow$29.12).
This matches the observation in~\cite{ugan2025pier} that training on monolingual data can hurt code-switched words.
A plausible explanation is that the model maps English words to phonetically similar Hindi words (Fig.~\ref{fig:example}).
The in-domain system (A5) reduces CS-WER and WER by similar relative amounts (23\% vs.\ 25\%), mainly through fewer substitutions and deletions. Exposure to real Hinglish dialogue thus appears necessary to recover CS words.

\noindent\textbf{Practical implications.}
Three observations may carry over to other low-resource clinical settings.
First, in-domain data mattered more than architecture. In-domain fine-tuning reduced DER by 0.8--1.1 points and tcpWER by 6.1 points. In contrast, the context network, the output size and 4$\times$ more out-of-domain pre-training data changed DER by at most 0.44 points.
Second, cascade modules should be evaluated together. Better diarization alone lowered extraction scores (Table~\ref{tab:synergy}).
Third, aggregate metrics can hide opposite trends on code-switched words. Generic Hindi fine-tuning (A1) changed WER by only $+0.5$ points, but increased CS-WER by $+0.9$ and CS substitutions by $+3.1$ points. We therefore recommend reporting point-of-interest metrics alongside WER for multilingual clinical ASR.

\vspace{-2mm}
\section{Conclusion}
\label{sec:conclusion}
\vspace{-1mm}

We presented DASHA, a modular cascade for extracting medical conditions from code-switched Hinglish doctor--patient conversations, which ranked first in the DISPLACE-M challenge.
With a w2v-BERT~2.0 EEND-VC diarization and a domain-adapted Qwen3-ASR, we reach 7.76\% DER and 19.27\% tcpWER using open models only (18.59\% with proprietary error correction), a new state of the art for two-speaker, code-switched Hindi SA-ASR.
Improvements in diarization and ASR are complementary for extraction.
Proprietary audio LLMs currently set the upper reference for extraction, and open omni-modal models remain far behind.
The diagnostic CS analysis suggests that the multilingual diarization encoder mainly reduces speaker confusion around code-switched words, and that generic Hindi fine-tuning can hurt their recognition.
Future work will use concept-level extraction metrics and extend the analysis to other language pairs.

\newpage

\section{Acknowledgments}
\label{sec:ack}
\sloppy This work was supported by the French National Research Agency (ANR-20-CE23-0012-01, MIM), the Agence de l'Innovation de D\'efense (2022~65~0079), and the EU Horizon 2020 project ELOQUENCE (101070558). Computations were performed on GENCI-IDRIS resources (grants AD011014044R2, A0191014044, AD011016519, AD011012177R5, A0181016176, AD011014274R2).
LLMs (Gemini~3 Pro, ChatGPT) were used only to edit the manuscript's wording; LLMs used as system components are described in Secs.~\ref{sec:system} and~\ref{sec:cs}.

\section{Compliance with Ethical Standards}
We use the de-identified DISPLACE-M corpus under the challenge's data agreement.
Diarization models are pre-trained on DIHARD3, VoxConverse, MSDWild, DISPLACE~2024, AMI, AliMeeting, AISHELL-4, NOTSOFAR-1 and RAMC.
ASR models are fine-tuned on FLEURS, IndicTTS, IndicVoices and VAANI.
All of these are existing public or licensed research corpora, used under their respective licenses and terms of use; no new human data were collected.
Proprietary LLMs were accessed only through enterprise API endpoints that do not retain data.

\let\oldthebibliography\thebibliography
\renewcommand{\thebibliography}[1]{\oldthebibliography{#1}\setlength{\itemsep}{0pt}\setlength{\parsep}{0pt}}
\bibliographystyle{IEEEbib}
\bibliography{refs_short}

\end{document}